\documentclass[reprint,superscriptaddress,amsmath,amssymb,aps,prd,twocolumn,floatfix,nofootinbib]{revtex4-2}
\usepackage{amsmath, amssymb}
\usepackage{listings}
\usepackage{graphicx}
\usepackage{dcolumn}
\usepackage{bm}
\usepackage{float,color}
\usepackage{xcolor}
\usepackage[normalem]{ulem}
\usepackage{tabularx}
\usepackage{mathtools} 
\usepackage{multirow}
\usepackage{amssymb}
\usepackage[caption=false]{subfig}
\usepackage{placeins}

\usepackage{scrextend}

\newcommand{\blue}{\color{blue}}

\usepackage[colorlinks=true,linkcolor=blue,citecolor=blue,urlcolor=blue]{hyperref}
\begin{document}
	
\title{Impact of neglecting center-of-mass acceleration in parameter estimation of stellar-mass black holes}
	
\author{Suvikranth Gera}
\affiliation{Department of Physics, IIT Guwahati, Guwahati, Assam, India}
\affiliation{School of Physical Sciences, Indian Association for the Cultivation of Science, Kolkata-700032, India}
\email{suvikranthg@gmail.com}
\author{Poulami Dutta Roy}
\affiliation{Chennai Mathematical Institute, Siruseri, 603103, India}
\affiliation{Department of Physics, University of Florida, PO Box 118440, Gainesville, Florida 32611-8440, USA}
\email{duttaroy.poulami@ufl.edu}
	
\begin{abstract}
A tertiary body near a coalescing binary can imprint its influence on the gravitational waves (GWs) emitted by that binary in the form of center-of-mass (CoM) acceleration. An example of such a scenario is a binary black hole (BBH) merging near a supermassive black hole, which is touted to occur frequently. The limited low-frequency sensitivity of current GW detectors makes it challenging to detect these effects, as the associated waveform phase remains elusive. However, next-generation (3G) detectors such as Cosmic Explorer (CE) and Einstein Telescope (ET), with improved sensitivity at lower frequencies, are expected to be capable of capturing such signatures. In our study, we focus on the stellar-mass BBHs and explore the parameter space where the CoM acceleration will play a dominant role affecting parameter inference of the binary. We demonstrate that an unaccounted CoM acceleration of a low-mass binary with a total mass of $5\, \rm{M}_{\odot} $ can lead to significant systematic biases, exceeding statistical errors in the estimation of the chirp mass and symmetric mass ratio when the CoM parameter $\alpha$ is as small as $\sim 10^{-9}$ and  $10^{-10}$ $\rm{s}^{-1}$ for CE and ET, respectively. We also find that asymmetric binaries are more susceptible to systematic bias when CoM acceleration is neglected. When the effect of CoM acceleration is included in the GW phase, then $\alpha = 10^{-7} \rm s^{-1}$ can be constrained with a bound of $10^{-9} (10^{-11})\, \rm s^{-1}$ for CE (ET). Our study thus highlights the crucial implications of considering the presence of a tertiary body in the GW emitted by a stellar-mass BBH when observed in 3G detectors.

\end{abstract}
	
\maketitle
	
\section{Introduction}

In the era of multimessenger astronomy, gravitational waves (GWs) from compact binaries have become a leading probe of gravity in the strong-field regime. The detection and parameter estimation techniques of these GWs often rely on a bank of waveform templates, which have traditionally been developed under the assumption that binary black hole (BBH) mergers occur in vacuum. This simplification removes the need to model external influences on the binary dynamics, allowing waveform construction to focus solely on intrinsic parameters such as the component masses and spins. The assumption is further justified by the sensitivity of the current LIGO–Virgo–KAGRA (LVK) network, which is primarily limited to the late inspiral and merger part of the GW signal or, equivalently, the higher frequency spectrum when expressed in the frequency domain. In this regime, the morphology of the waveform is predominantly dictated by the binary dynamics. 

In reality, BBHs are expected to form and evolve in a variety of dense environments such as gas-rich active galactic nuclei (AGN) disks \cite{Yang:2020lhq,Bartos:2016dgn,OLeary:2008myb}, globular clusters \cite{Rodriguez:2016kxx,Tiwari:2023cpa}, and nuclear star clusters \cite{Mapelli:2021gyv}. Numerical studies suggest that the cosmological BBH merger rate in (or near) galactic nuclei can be at the level of a few Gpc$^{-3}$ yr$^{-1}$ \cite{Laeuger:2023qyz, Yang:2020lhq, Bartos:2016dgn, OLeary:2008myb, Rodriguez:2016kxx}. In particular, a more predominant configuration in such scenarios would be that of a stellar-mass BBH orbiting a supermassive black hole (SMBH) at a galactic center. In these hierarchical triple systems, the SMBH acts as a perturbation, inducing a center-of-mass (CoM) acceleration on the binary. The magnitude of this acceleration depends primarily on the  BBH–SMBH separation and the mass of the SMBH \cite{Laeuger:2023qyz,Antonini:2012ad}. Such systems have gained renewed prominence in recent years, as hierarchical mergers are increasingly considered a plausible formation channel for intermediate mass black holes \cite{Rantala:2024crf,Liu:2023zea, Fujii:2024uon}. Notably, recent event such as GW190814\cite{LIGOScientific:2020zkf} is hypothesized to result from the final stage merger in a hierarchical triple system involving a tertiary 23$M_\odot$ black hole and a pair of neutron stars, which may have previously merged to form a 2.5$M_\odot$ compact object that subsequently coalesced with the tertiary.

Beyond the CoM acceleration, other environmental phenomena include gas drag in AGN disks and dynamical friction in dense stellar clusters experienced by the binary. Collectively, the influences arising from the non-vacuum surrounding of the binary are often referred to as ``\textit{environmental effects}''. Such environmental effects modify the GW signal through additional forces or accelerations, with the most prominent signatures expected in the lower frequency spectrum of the GW signal\cite{Barausse:2014tra,Barausse:2014pra,Barausse:2007dy,Cardoso:2019rou,Sberna:2022qbn,Cole:2022ucw,Zwick:2024yzh,Roy:2024rhe,CanevaSantoro:2023aol,Chen:2020lpq,Caputo:2020irr,Coogan:2021uqv,Vijaykumar:2023tjg,DuttaRoy:2025gnu}.

The current LVK network, with its limited power spectral density at lower frequencies, is not suited for detecting these environmental signatures. Dedicated searches for environmental imprints in existing data have so far found no compelling evidence. For example, the binary neutron star event GW170817 yields the most stringent upper limit on ambient density, effectively ruling out very dense gas in that case \cite{CanevaSantoro:2023aol}. In contrast, third-generation (3G) ground-based detectors like Cosmic Explorer (CE) and Einstein Telescope (ET), with their improved low-frequency sensitivity will, in principle, be capable of measuring these effects. This prospect has renewed interest in understanding and incorporating these effects into waveform modelling. To fully exploit the potential of  3G detectors, the waveform models must be extended beyond the vacuum approximation, failing which could risk introducing systematic biases in future parameter inference \cite{DuttaRoy:2025gnu,Garg:2024qxq,DeLuca:2025bph}. Conversely, a confident detection of environmental effects would open a new window on probing AGN disks, dense stellar systems, and dark-matter structures.

Focussing on CoM acceleration, recently multiple works of literature have examined its effects on binary systems. For example, in the context of space-based detector LISA, several works have investigated the detectable imprints of different  CoM acceleration sources on various compact binaries \cite{Yunes:2010sm,Ebadi:2024oaq,Tamanini:2018cqb}. On the waveform modeling front, a consistent numerical framework for incorporating CoM acceleration as corrections to non-spinning post-Newtonian (PN) order waveforms was first presented in \cite{Vijaykumar:2023tjg}, with an independent extension to spinning compact objects carried out by Lazarow et al. \cite{Lazarow:2024gdn}.

Apart from the implications already studied, various interesting phenomena remain unexplored. For example, one would expect effects such as  Kozai–Lidov oscillations to excite large eccentricities depending on the mutual inclination of the orbits \cite{Kozai:1962zz,Lidov:1962wjn}. Also, additional imprints can arise from gravitational lensing by the tertiary \cite{DOrazio:2019fbq} and Shapiro time delay \cite{Shapiro:1964uw}.

Motivated by this, we focus on hierarchical triples in which a stellar-mass BBH orbits a massive tertiary (e.g., an SMBH). We quantify how CoM acceleration from the tertiary imprints on the GW phase and assess the consequences for parameter estimation.  Our analysis places special emphasis on low-mass stellar binaries, highlighting that neglecting CoM acceleration can introduce significant systematic biases that surpass the statistical uncertainties in parameter inference. The chirp mass and symmetric mass ratio are especially vulnerable to these biases, which could severely compromise the accuracy of astrophysical interpretations. We also identify that asymmetric binaries exhibit an enhanced susceptibility to these systematic effects, amplifying the necessity of incorporating environmental considerations into GW waveform modeling.

Our results have several implications. Detecting and characterizing CoM acceleration would directly probe the environments of compact-object mergers and help distinguish isolated formation channel from dynamically assembled binaries and the potential growth of intermediate-mass black holes via hierarchical mergers in dense environments \cite{Zwick:2025ine,Inayoshi:2017hgw}. More accurate modeling will also reduce systematic biases in population studies and inferences of fundamental source properties. Finally, as 3G detectors extend access to lower frequencies, CoM acceleration and related phenomena will broaden the scope of GW astronomy for tests of fundamental physics in extreme astrophysical settings.

This paper is organized as follows. In Sec.~II we outline the derivation of the contribution of CoM acceleration to the binary’s GW phase which was studied in \cite{Lazarow:2024gdn} and outline the Fisher-matrix formalism used to obtain bounds on the parameters, including the CoM-acceleration coefficient. In Sec.~III we present our main results: (i) the systematic biases in chirp mass and mass ratio when CoM acceleration is neglected, and (ii) measurability forecasts for CE and ET across representative binary configurations. We summarize the main conclusions in Sec.~IV.

\section{Analysis}

\subsection{Waveform Modification Due to Center-of-Mass Acceleration}
\begin{figure*}[t]
    \centering
    \subfloat[]{
  \includegraphics[width=0.41\textwidth]{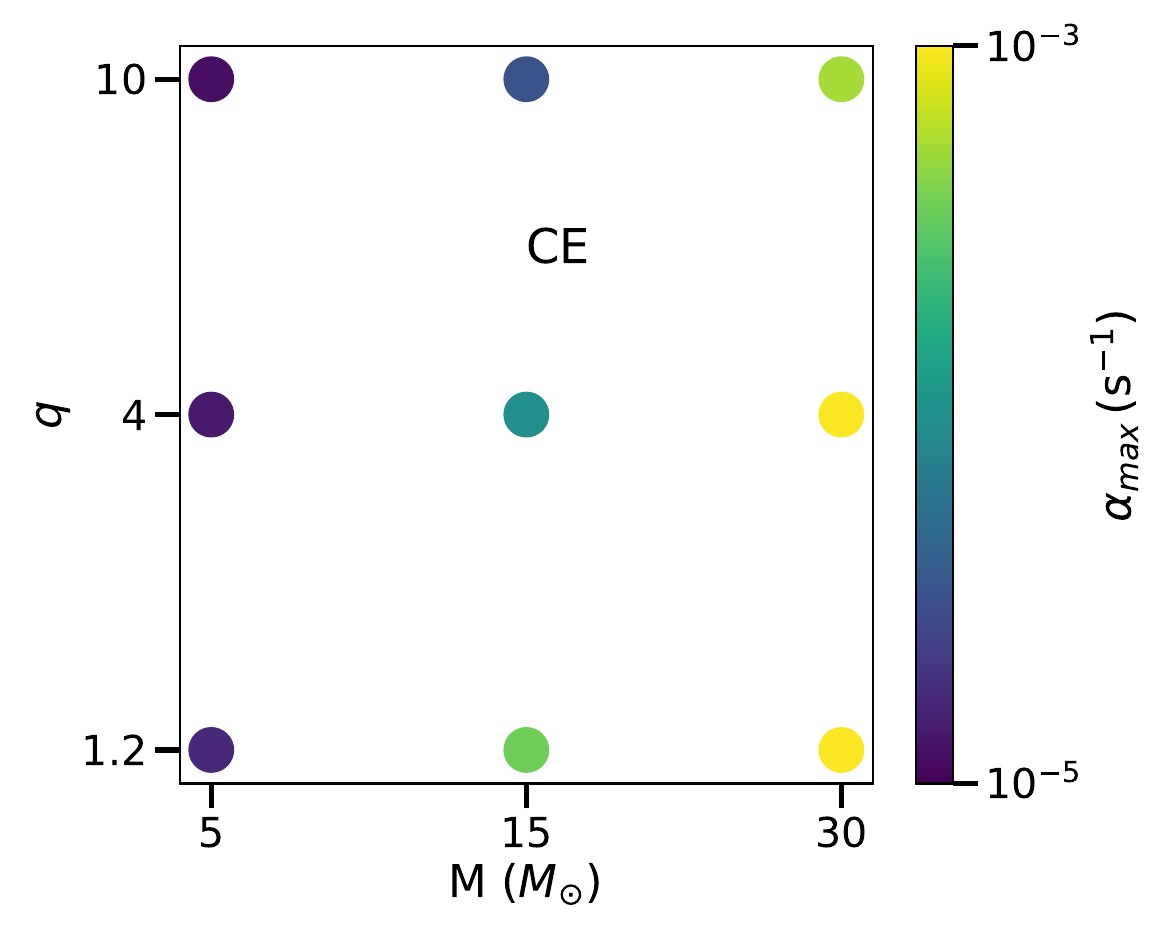}%
    }
    \hspace{0.3in}
   \subfloat[]{
  \includegraphics[width=0.41\textwidth]{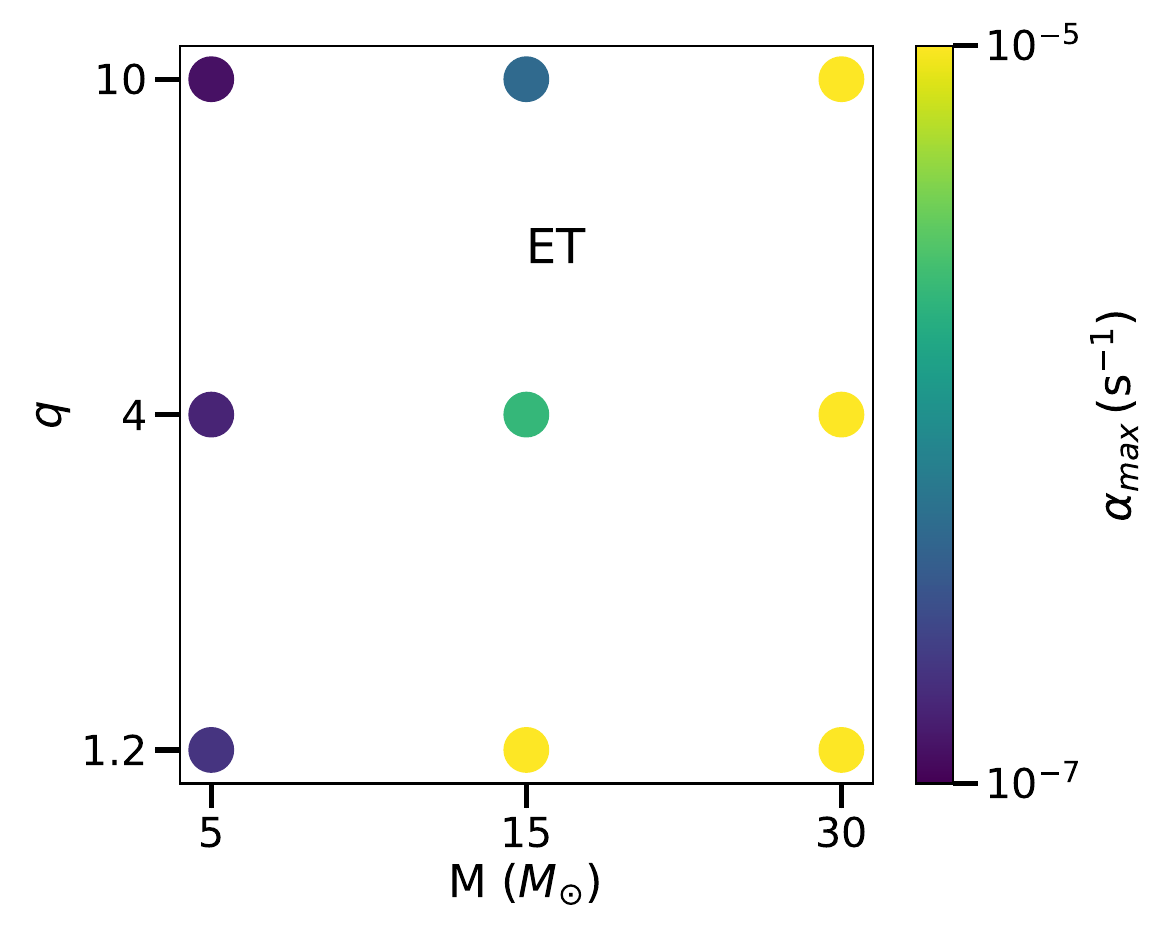}%
} 
    \caption{The plot shows the allowed upper value of $\alpha$ i.e. $\alpha_{max}$ for different binaries considered in our study when observed through CE and ET. Note that $\alpha_{max}$ is an order of magnitude lower for a particular binary in ET in comparison to CE.}
    \label{fig:alpha_range}
\end{figure*}	
In the present work, we consider a three-body system in which a binary is undergoing coalescence while orbiting a third compact object. We assume that the third body is sufficiently far from the binary such that the coalescence dynamics of the binary remain quasi-circular and effectively independent of the third body's influence.
	
However, the presence of the third body induces an acceleration on the binary’s center of mass (CoM), which manifests as a time-varying Doppler shift in the phase of the gravitational wave signal (generated from the binary coalescence). This effect is implemented in the waveform  by reparameterization of the time t by $t \rightarrow t + \alpha(t - T)$ to first order in the CoM acceleration, where $ \alpha $ quantifies the CoM acceleration and $ T $ is a reference time.
	
This formalism is developed in \cite{Lazarow:2024gdn}, where the authors considered the stationary phase approximation (SPA) of the binary coalescence waveform to compute the correction terms contributed due to CoM acceleration upto 3PN in the `Taylor' model of waveforms. We include these correction terms to the already existing circular and spin contributions upto 4.5PN. The waveform can be written as

\begin{eqnarray}
    \Psi (f)& =& \phi_c + 2 \, \pi \,f \, t_c  +  \frac{3}{128 \, \eta \,v^5} \Big( 1 + \Delta \Psi_{4.5 \mathrm{PN}}^{\mathrm{pp,circ.}} + \nonumber\\ && \Delta \Psi_{4 \mathrm{PN}}^{\mathrm{spin,circ.}} 
    + \Delta \Psi_{3 \mathrm{PN}}^{\mathrm{CoM}}\Big) \,
    \label{eq:PNphase}
\end{eqnarray}

where $ v = (\pi M f)^{1/3} $ is the orbital velocity parameter and $M$ being the total mass of the binary. The coalescence time is denoted by $ t_c $ and $ \phi_c $ is the coalescence phase. 
The term $\Delta \Psi_{4.5 \mathrm{PN}}^{\mathrm{pp,circ.}}$ represents the circular point particle contribution to the phase extended till 4.5PN ~\cite{Blanchet:2023bwj}
\begin{equation}
\Delta \Psi^{\mathrm{pp, circ.}}_{4.5 \mathrm{PN}}(f)= \sum_{k=0}^{9}\left(\phi_k\,v^k+ \right.
    \phi_{kl} v^k \ln v  + 
    \left.\phi_{kl2} v^k \ln^2 v\right) 
   \label{eq:PNphase_circ}
\end{equation}
where $\phi_k$ are the PN coefficients which are functions of source parameters \cite{Buonanno:2009zt,Blanchet:2023bwj}. The spin contribution to the circular part  $\Delta \Psi_{4 \mathrm{PN}}^{\mathrm{spin,circ.}}$ upto 4PN can be found in~\cite{Arun:2004hn, Arun:2008kb,Buonanno:2009zt,Mishra:2016whh}.
The correction due to CoM acceleration upto 3PN are of the form
\begin{equation}
    \Delta \Psi_{3 \mathrm{PN}}^{\mathrm{CoM}} = \alpha v^{-5} \sum_{n=0}^6 A_n v^{-8+n}
    \label{eq:CoM_phase}
\end{equation}
where the coefficients $A_n$ are functions of total mass, mass ratio and spins of the binary components. The exact expressions of $A_n$ can be found in the appendix of \cite{Lazarow:2024gdn}. The parameter $\alpha$ denotes the magnitude of the CoM acceleration which is determined by the mass of the tertiary, $M_3$ (in our case the SMBH mass) and the radial distance from it (see eq.2b in \cite{Lazarow:2024gdn})
\begin{equation}
    \alpha = 9.9 \times 10^{-12} \rm s^{-1} \frac{M_3}{M_{\odot}} \Big(\frac{1 \rm AU}{r} \Big)^2
    \label{eq:alpha}
\end{equation}
The GW strain in frequency domain can hence be written as
\begin{align}
    \Tilde{h}(f) = \mathcal{A}(f) \, e^{i \Psi (f)} = \hat{\mathcal{A}} f^{-7/6} e^{i \Psi(f)} \,,
    \label{eq:waveform}
\end{align}
where 
\begin{equation}
    \hat{\mathcal{A}} = \frac{1}{\sqrt{30} \pi^{2/3}} \frac{M_c^{5/6}}{d_L} \,.
\end{equation}
The $M_c = (m_1 m_2)^{3/5}/M^{1/5} = M\eta^{3/5}$ is the chirp mass, $M=m_1+m_2$ is the total mass of the source, $m_1$ and $m_2$ are the masses of primary and secondary component of the binary, $\eta$ is the symmetric mass ratio, and $d_L$ is the luminosity distance to the source. The phase $\Psi(f)$ has the form denoted in eq.(\ref{eq:PNphase}). The detector-frame total mass is related to the source-frame total mass ($M_{\rm s}$) as $ M = (1+z) M_{\rm s}$, 
where $z$ is the redshift to the source. Considering flat Lambda-CDM cosmology, $d_L$ and $z$ are related by
\begin{equation}
d_{L} (z) = \frac{(1+z) }{H_0} \int_0^z \frac{dz'}{\sqrt{\Omega_M (1+z')^3 + \Omega_\Lambda}} \,,
\label{eq:redshift}
\end{equation}
where the cosmological parameters are $\Omega_{M} = 0.3065$, $\Omega_\Lambda = 0.6935$ and $h = 0.6790$ with $H_0 =100h$ (km/s)/Mpc~\cite{Planck:2018vyg}.

Before moving on with the Fisher analysis, a few comments are in order. We would like to emphasize that a similar formalism was carried out by \cite{Vijaykumar:2023tjg} to account for CoM corrections for the spinless case. The amplitude corrections have also been ignored in our study as has been done \cite{Lazarow:2024gdn}. Note that these corrections are first order in CoM acceleration as discussed above. This sets the upper bound on  $\alpha$ before higher order terms become relevant as given by the relation
\begin{equation}
    \alpha \ll \frac{\partial \Psi/\partial \alpha}{\partial^2 \Psi/\partial \alpha^2}
\end{equation}

where the RHS is computed using the phase expression of eq.\eqref{eq:PNphase}. As the CoM corrections arise at negative PN order, the upper bound on $\alpha$ is obtained at the cutoff frequency of the detector defined by $f= f_{low}$. Hence, the bound $\alpha_{max}$ is a function of $\alpha_{max}(f_{low},M,q)$ where $q=m_1/m_2$. We consider the cutoff frequency $f_{low}= \rm 5 Hz (1 Hz)$ for CE (ET). The maximum allowed values of $\alpha$ for different binary configurations are shown in Fig.(\ref{fig:alpha_range}).

Due to the superior low frequency sensitivity of ET we observe that the magnitude of $\alpha_{max}$ for ET is at least an order lower than CE. This is expected as a better sensitivity would result in higher order corrections in $\alpha$ becoming more relevant at much lower frequencies. The value of $\alpha_{max}$ is smallest for the low-mass binary, being of the order of $10^{-5} (10^{-7} )\, \rm s^{-1}$ for CE (ET). As in the relationship with total mass, we observe that $\alpha_{max}$ increases with increasing total mass, assuming that the mass ratio is the same, while it reduces with increasing mass ratio assuming that total mass remains fixed.

\begin{figure*}[tb]
    \centering
    \subfloat[]{
  \includegraphics[width=0.41\textwidth]{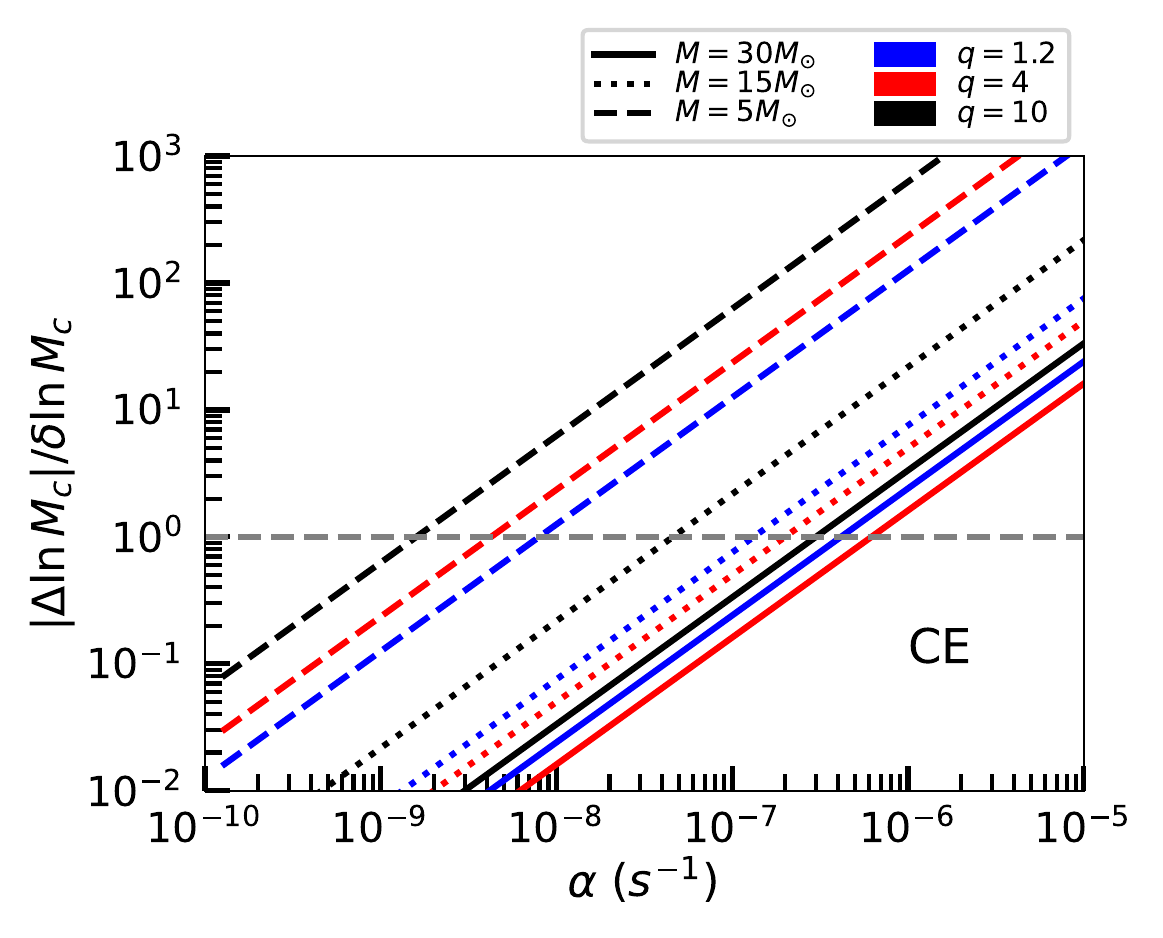}%
    }
    \hspace{0.3in}
   \subfloat[]{
  \includegraphics[width=0.41\textwidth]{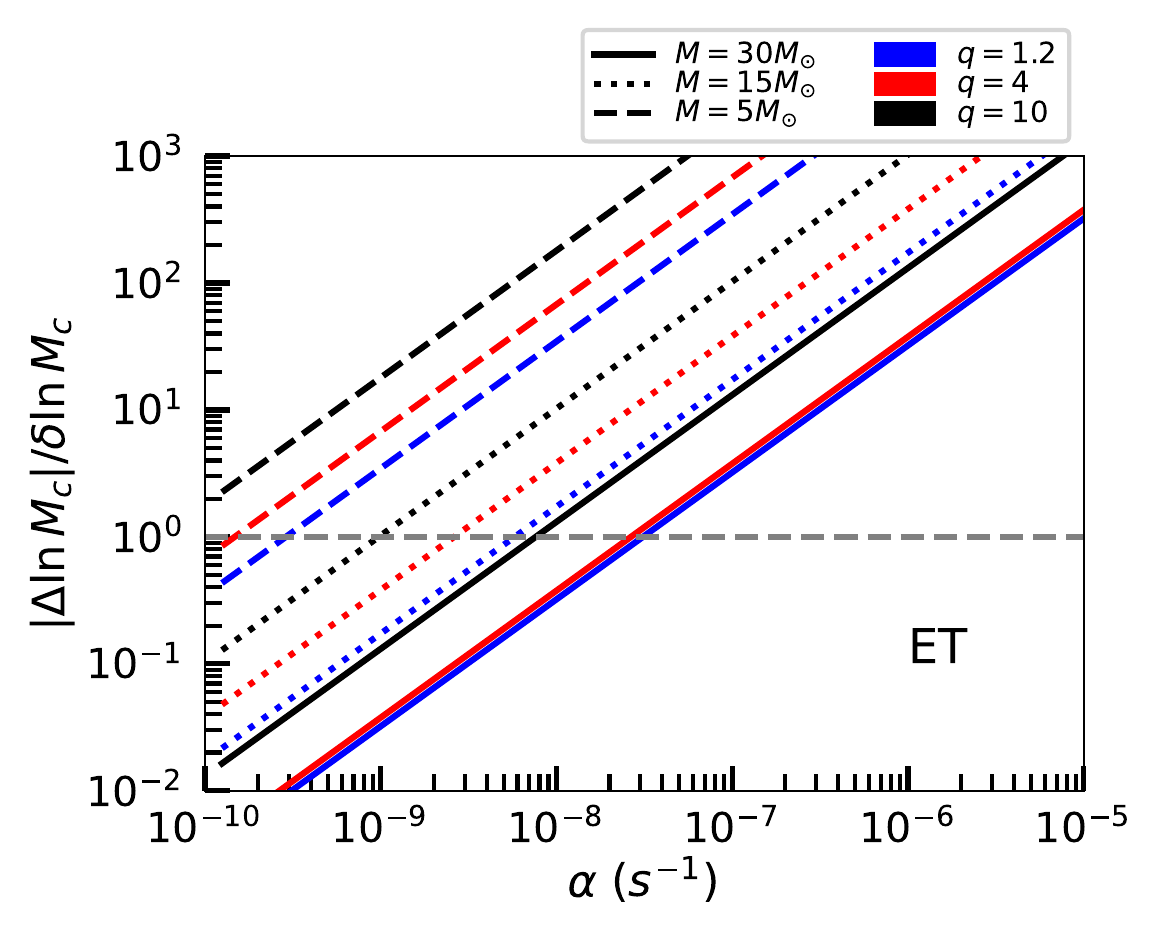}%
} 
    \caption{The plot shows ratio of absolute value of systematic bias over statistical one for $\rm ln M_c$ for different values of $\alpha$. Binary black holes of different mass and mass ratio are considered with aligned spin magnitudes $(0.2, 0.1)$ at a distance of 500 Mpc. The systematic error dominates over statistical for lowest value of $\alpha$ being $\sim 10^{-9} \rm s^{-1}$ for CE and $\sim 10^{-10} \rm s^{-1}$ for ET.}
    \label{fig:Mc_bias}
\end{figure*}

\subsection{Parameter Estimation Using the Fisher Information Matrix}

Fisher information matrix(FIM) provides an analytic framework to obtain bounds on model parameters.
In practice, the Cramer-Rao theorem establishes that the inverse of the FIM provides the lower bounds on the variance of the parameters. In the context of GW physics, the FIM was first systematically applied to compact binaries by Cutler and Flanagan \cite{Cutler:1994ys} and later developed further by Poisson and Will \cite{Poisson:1995ef}, where it quickly became a standard method for parameter estimation. Under the high signal-to-noise ratio(SNR) approximation, with stationary Gaussian noise, we can write the probability distribution of waveform parameters $\bm \theta$, given data $d(t)$ and best-fit parameters $\bm{\hat{\theta}} $, maximizing the Gaussian likelihood as

\begin{equation}
 p({\bm \theta}|d) \propto p^{0}({\bm \theta}) \exp\left[ -\frac{1}{2} \Gamma_{ab} (\theta_{a} - \hat{\theta}_{a}) (\theta_{b} - \hat{\theta}_{b}) \right]
\end{equation}
where $ \Gamma_{ab} $ is the FIM, defined by
\begin{eqnarray} \label{eq:fisher}
	\Gamma_{ab} &=& \left( h_{,a}  \mid h_{,b}  \right)\Big|_{\bm{\theta} = \bm{\hat{\theta}}} \\ \nonumber
   &=& 2 \int_{f_{\text{low}}}^{f_{\text{up}}} \frac{\tilde{h}_{,a}^*(f)\tilde{h}_{,b}(f) + \tilde{h}_{,b}^*(f)\tilde{h}_{,a}(f)}{S_n(f)} df
\end{eqnarray}
with $ \tilde{h}_{,a}(f) \equiv \partial \tilde{h}(f)/\partial \theta^a $. The tilde denotes the Fourier transform of the time-domain signal $h(t)$ and `$*$' denotes the complex conjugate. In the absence of any bias, the `best-fit' values of the parameters will correspond to their true values. The SNR $\rho$ for a signal $h(t)$ is defined as
\begin{eqnarray}
    \rho^2 = 4 \int_{f_{\rm low}}^{f_{\rm up}} \frac{|\tilde{h}(f)|^2}{S_n (f)} df \,,
\end{eqnarray}
where $f_{\rm low}$ and $f_{\rm up}$ are the lower and upper cut-off frequencies that depend on the detector sensitivity and properties of the source.
	
The inverse of the Fisher matrix provides the \textit{variance-covariance matrix} ($\Sigma^{ab}$) whose diagonal elements represent the 1$\sigma$ statistical uncertainty for each parameter \cite{Poisson:1995ef}
\begin{equation}
	\Sigma^{ab} = (\Gamma^{-1})^{ab}, \, \sigma_a = \sqrt{\Sigma_{aa}}
\end{equation}

\begin{figure*}[t]
    \centering
    \subfloat[]{
  \includegraphics[width=0.41\textwidth]{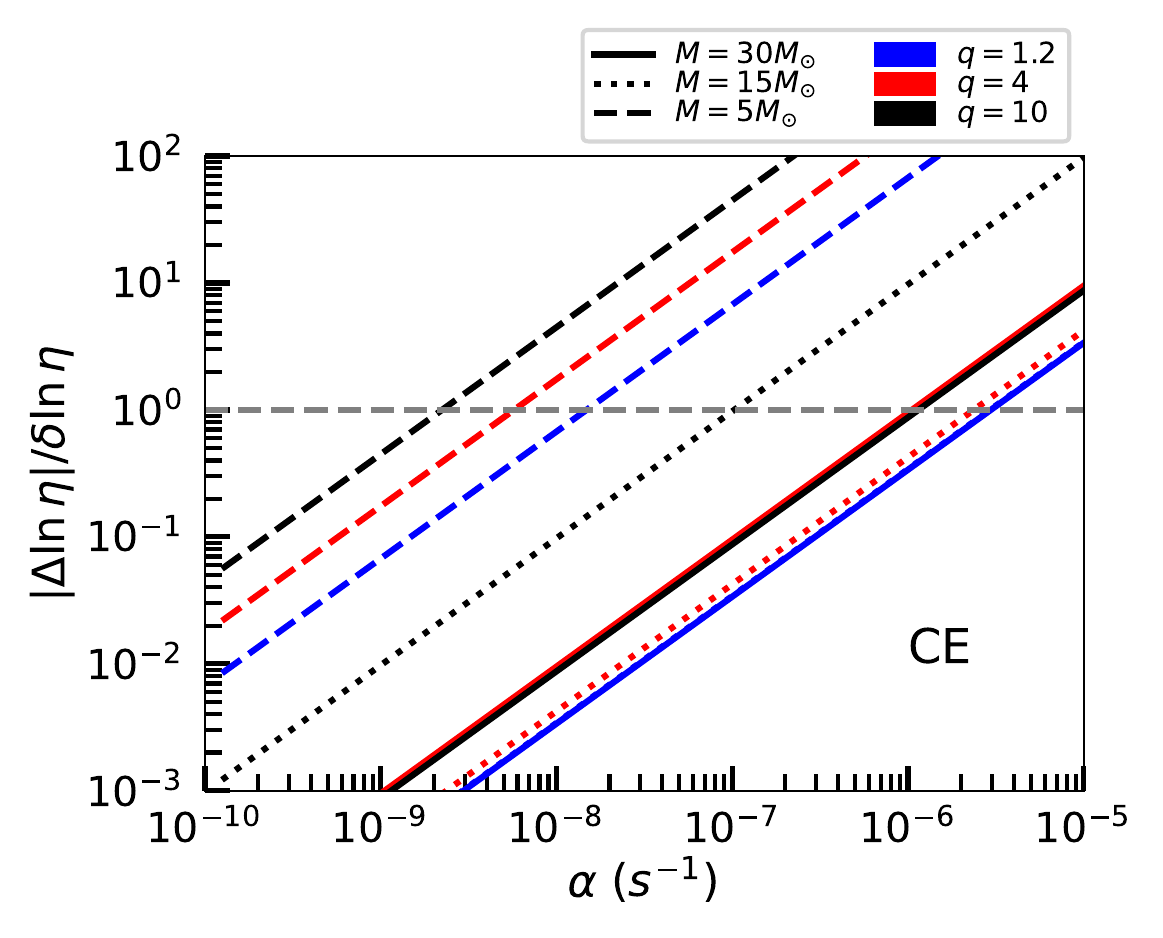}%
    }
    \hspace{0.3in}
   \subfloat[]{
  \includegraphics[width=0.41\textwidth]{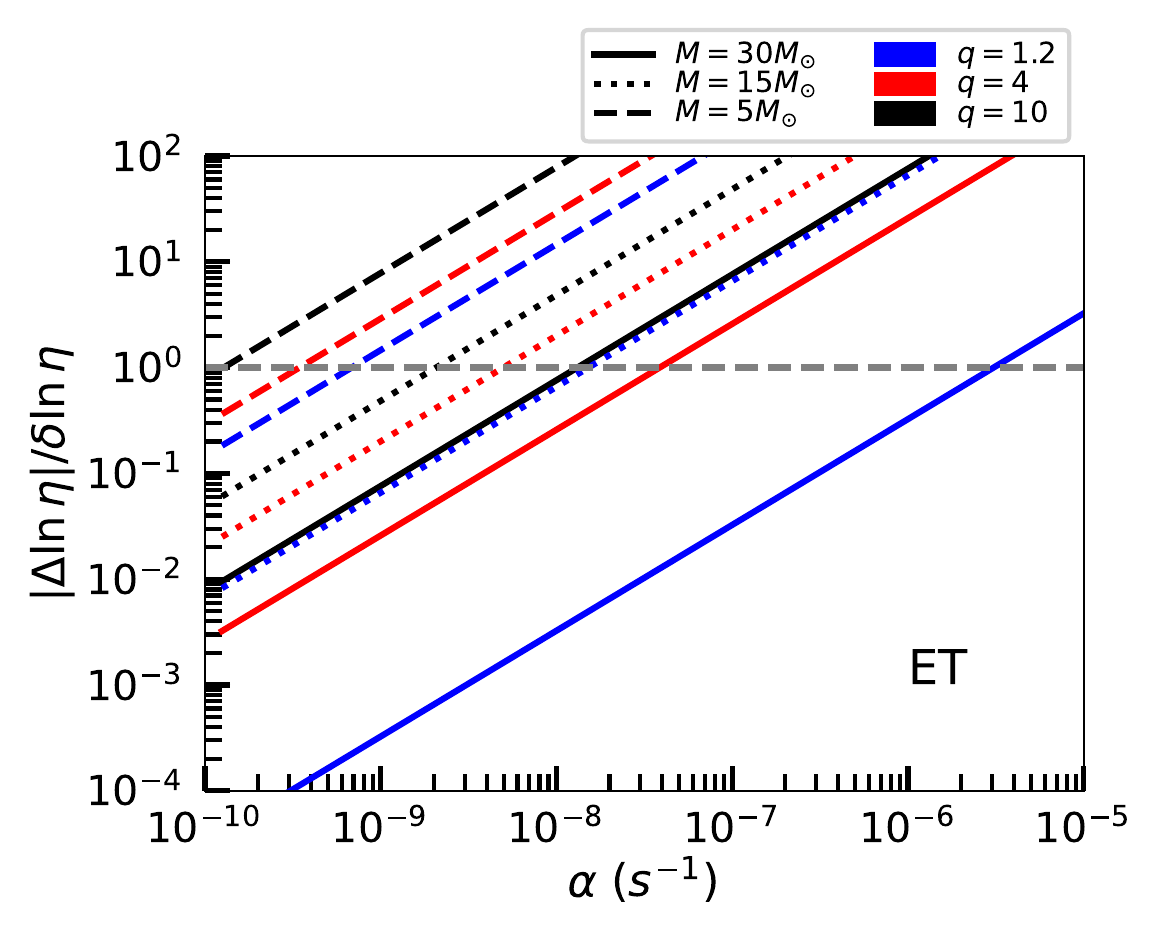}%
} 
    \caption{The plot shows ratio of absolute value of systematic bias over statistical one for $\rm ln \eta$ for different values of $\alpha$. Binary black holes of different mass and mass ratio are considered with aligned spin magnitudes $(0.2, 0.1)$ at a distance of 500 Mpc. The systematic error dominates over statistical for lowest value of $\alpha$ being $\sim 10^{-9} \rm s^{-1}$ for CE and $\sim 10^{-10} \rm s^{-1}$ for ET.}
    \label{fig:eta_bias}
\end{figure*}

Apart from the statistical errors, which are determined by the variance-covariance matrix as discussed above, the waveform models themselves are susceptible to modeling errors that further induce errors in the parameter estimation. These errors are not statistical in nature and are known as systematic errors. The systematic errors can be of various factors and range in their magnitudes; however, they become statistically relevant in parameter estimation when their magnitude is of the same order or higher than statistical measurement uncertainety.
    
To determine these systematic errors, we follow the Cutler-Vallisneri formalism \cite{CutlerVallisneri07}. In this approach, we assume the `true model' ($\tilde{h}_{T}$) accurately describing the GW and an approximate model ($\tilde{h}_{AP}$) defined as 
	\begin{equation}
		\tilde{h}_{AP}=\mathcal{A}_{AP} e^{i \Psi_{AP}}
	\end{equation}
while assuming that the true waveform $\tilde{h}_{T}$ differs from $\tilde{h}_{AP}$ in amplitude and phase by $\Delta\mathcal{A}$and $\Delta \Psi$ giving
	\begin{equation}
		\tilde{h}_T =\left( \mathcal{A}_{AP}+\Delta\mathcal{A}\right)e^{i\left[\Psi_{AP}+\Delta\Psi\right]}
	\end{equation}
	
In our case, the systematic error is the difference in the true parameter value ($\theta_a^{\rm T}$) when the CoM acceleration effect is considered compared to the best-fit value ($\hat{\theta}_a$) estimated while neglecting CoM acceleration. This is quantified by
	\begin{equation}
		\Delta \theta^a = \theta_a^{\rm T} - \hat{\theta}_a\,\approx \Sigma^{ab}\left((\Delta\mathcal{A}+i\mathcal{A}_{AP}\Delta\Psi)e^{\Psi_{AP}}\mid \partial_b \tilde{h}_{AP}\right)
	\end{equation}
Since we consider no contribution to the amplitude due to CoM acceleration, $\Delta \mathcal{A} = 0$. While calculating the systematic error, we consider the approximate waveform to be that of the quasi-circular contribution and the $\Delta\Psi = \Delta \Psi^{\rm CoM}_{\rm 3PN}$ (see eq.(\ref{eq:CoM_phase})).

We consider the parameter space of $\bm{\theta}^a = (t_c,\phi_c, \ln M_c,\ln \eta, \chi_1,\chi_2, \ln\alpha, \ln d_L)$, where $\chi_{1,2}$ denotes the dimensionless spin magnitudes along the orbital angular momentum. The upper limit $f_{up}$ is set to be the frequency corresponding to the innermost stable circular orbit (ISCO) of the remnant which we assume to be a Kerr BH. The complete expression for the Kerr ISCO can be found in Appendix C of \cite{Favata:2021vhw}. The noise PSD for CE and ET are taken from \cite{Kastha:2018bcr} and \cite{Hild:2010id} respectively. Note that to take account of the triangular shape of ET, we include a factor of $\sqrt{3}/2$ in the amplitude of the waveform.
 	
In this study, we consider the BBHs to occur at a luminosity distance of $d_L= 500 \rm Mpc$ with an aligned spin of $(\chi_1=0.2,\chi_2=0.1)$. As representatives of stellar-mass BBH, we consider three different systems with $M=\left(5 M_{\odot}, 15 M_{\odot},30 M_{\odot} \right) $ with mass ratios $q= \left(1.2,4,10\right)$.  The SNR of the binaries considered are of $\sim\mathcal O(100)$ for both  CE and ET.

\section{Results}
	
\subsection{Systematics on binary parameters}

We now present our results for the computation of the systematic bias in the measurement of binary parameters when the CoM acceleration in neglected for different systems. To this extent, we focus on the measurement of chirp mass and symmetric mass ratio. In order to highlight the importance of systematic error, we plot the ratio of the absolute value of systematic bias to statistical bias i.e. $\Delta \theta/ \delta\theta$, which when crosses unity signifies the systematic error dominating the statistical uncertainity. In this regime of parameter values, the CoM acceleration plays a crucial role, and if neglected, will severely bias the parameter estimations.
 
\begin{figure*}[t]
    \centering
    \subfloat[]{\label{fig:bound}%
  \includegraphics[width=0.41\textwidth]{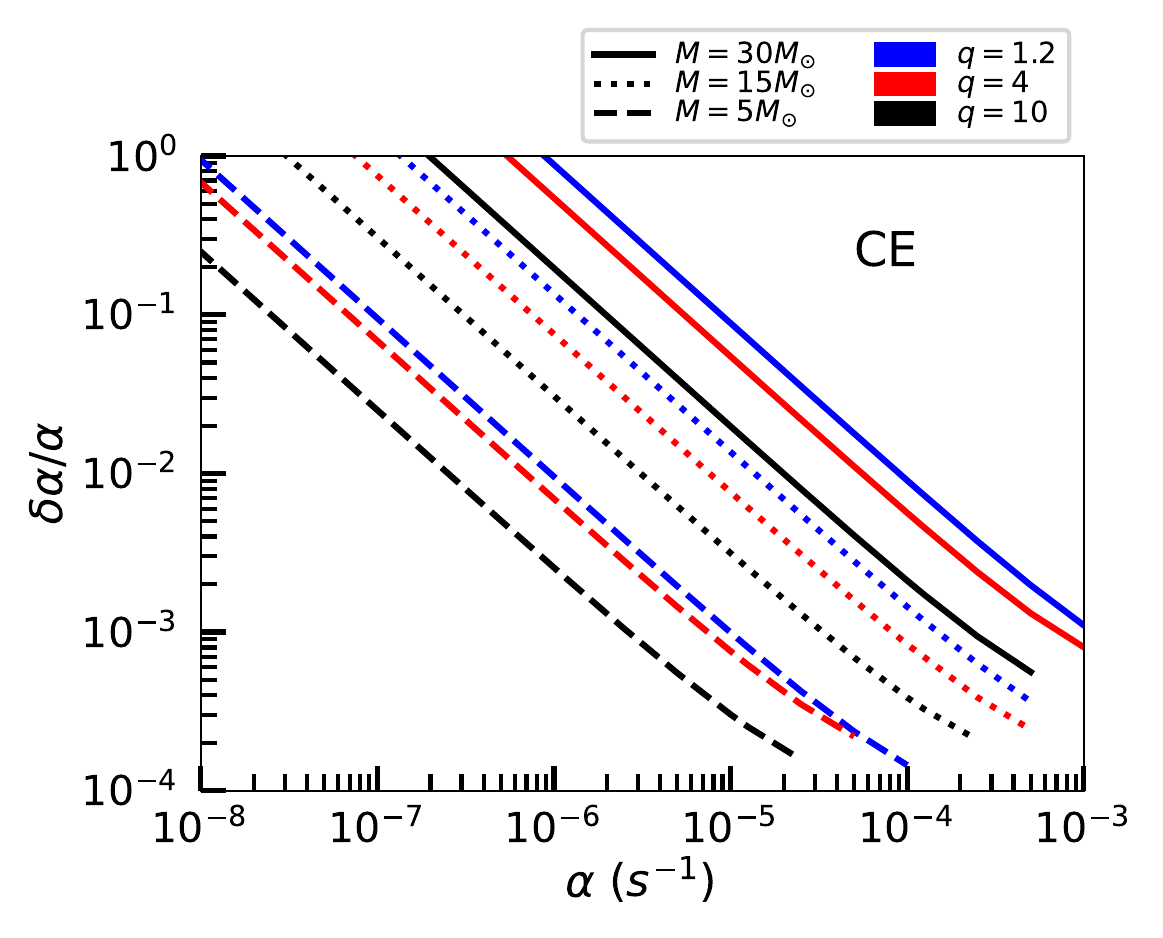}%
    }
    \hspace{0.3in}
   \subfloat[]{\label{fig:bound_q}%
  \includegraphics[width=0.41\textwidth]{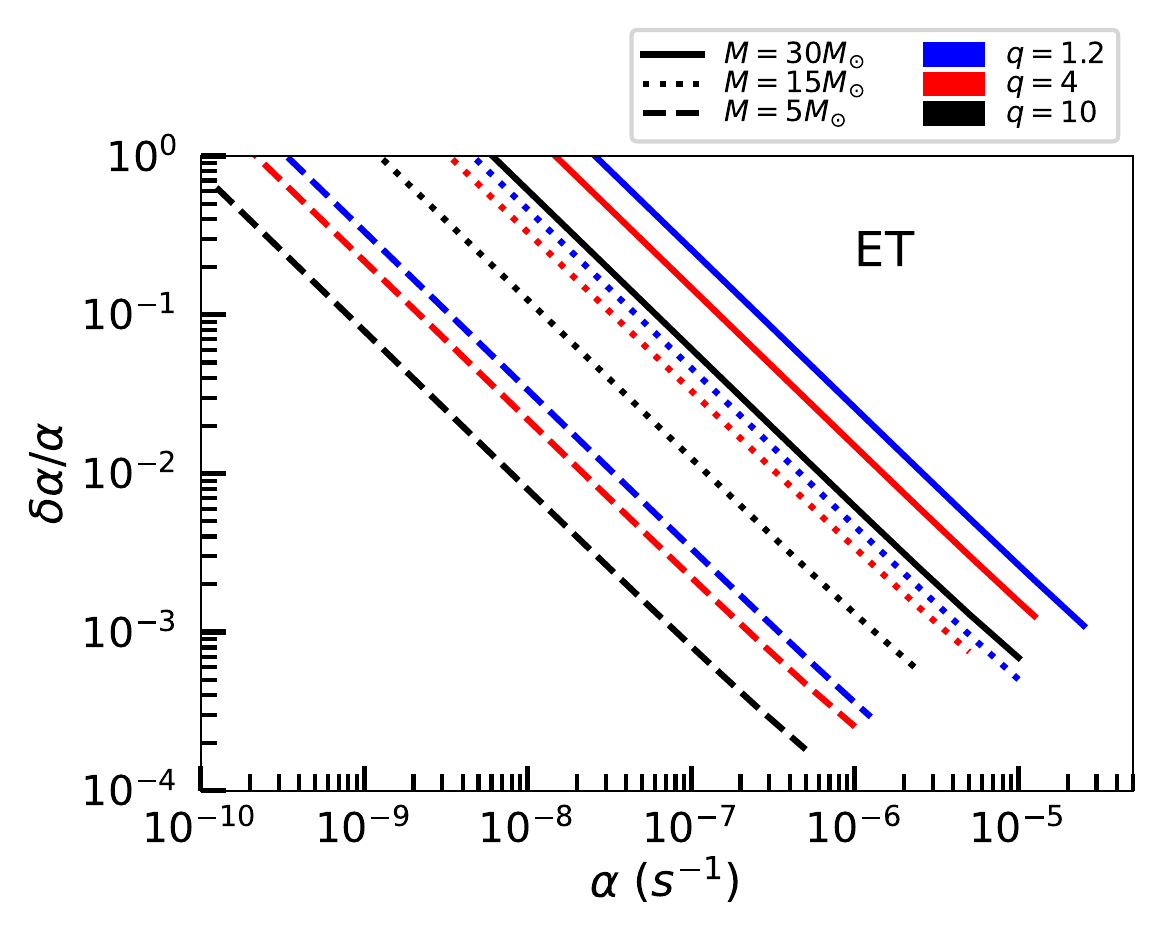}%
} 
    \caption{The plot shows the fractional error on $\rm ln \alpha$ for different values of $\alpha$. Binary black holes of different mass and mass ratio are considered with aligned spin magnitudes $(0.2, 0.1)$ at a distance of 500 Mpc.}
    \label{fig:alpha_bound}
\end{figure*}

The main results of the paper are shown in Fig.\ref{fig:Mc_bias}
and Fig.\ref{fig:eta_bias} which shows the variation of magnitude of systematic over statistical error for different binaries as a function of the CoM acceleration parameter $\alpha$. Note that in both cases, the value of $\alpha$ is lower for ET than CE for which the systematic error dominates over statistical bias. This is expected due to the slightly better low-frequency sensitivity of ET where the effects of the CoM acceleration are expected to dominate. Also, the parameter estimation of the low-mass binaries is severely affected when the CoM acceleration is neglected. In our study, we have considered $5 \rm M_{\odot}$ as the lowest mass BBH and, for all the mass ratios considered, we observe the systematic error dominating the statistical bias for much lower values of $\alpha$. The effect of CoM acceleration is critical for such low-mass systems due to their long inspiral with multiple cycles. The systematic bias accumulates over such a long inspiral and becomes a serious cause of concern even for very small values of $\alpha$. We also note that systematic bias is more prominent for asymmetric binaries. The coefficients in eq.(\ref{eq:CoM_phase}) vary inversely with $\eta^2$ and more asymmetric binaries have smaller $\eta$ values, making the CoM acceleration effect dominant, leading to the trend observed in our results. It is important to keep in mind that the allowed maximum value of $\alpha$ is different for each binary configuration as shown in Fig.\ref{fig:alpha_range}. 
 
Focusing first on chirp mass, the statistical error with and without CoM acceleration effect in the phase is $\mathcal{O}(10^{-4}) $ and $\mathcal{O}(10^{-5}) $ respectively and the systematic bias is negative for all considered parameters. The systematic error dominates over statistical for lowest value of $\alpha$ being $\sim 10^{-9} \rm s^{-1}$ for CE and $\sim 10^{-10} \rm s^{-1}$ for ET. For heavier BBHs, in case of CE, we observe the mass ratio $q=1.2$ with systematic bias dominating over statistical error at lower $\alpha$ in comparison to $q=4$. This can be attributed to the interplay between the number of cycles spent in the detector sensitivity band, the strength of the CoM acceleration dephasing along with the sensitivity of the detector in those frequency regimes.
We observe similar trends in the case of estimation of symmetric mass ratio with the lowest value of $\alpha$ being $\sim 10^{-9} \rm s^{-1}$ for CE and $\sim 10^{-10} \rm s^{-1}$ for ET for which the systematic bias dominates over the statistical error. A similar observation has been made for binary neutron stars with zero spins in the Appendix of \cite{Lazarow:2024gdn}, where the chirp mass and symmetric mass ratio have been shown to deviate significantly from their true values for different $\alpha$. 

\subsection{Measurement of CoM acceleration}

In this section, we incorporate the dephasing due to CoM acceleration in the GW phase and compute the bounds on the parameter $\alpha$ for different binaries. Fig.\ref{fig:alpha_bound} shows the variation of fractional error on $\alpha$ as a function of $\alpha$ for CE and ET. Note that the curves for different binaries end at different points corresponding to the value of allowed $\alpha_{max}$ for that particular system (see Fig.\ref{fig:alpha_range}). The fractional error reduces with increasing $\alpha$ as expected. As is evident from Fig.\ref{fig:alpha_bound}, a better constraint on $\alpha$ is obtained from the ET observation. For example, an $\alpha = 5 \times 10^{-7} \rm s^{-1}$ is measured with an accuracy of $9 \times 10^{-11} s^{-1}$ for a binary of total mass $5 M_{\odot}$ and $q=10$ when observed in ET. Following \cite{Lazarow:2024gdn}, we assume the detectability criterion for CoM acceleration effect to be $\delta\alpha/\alpha < 1$, then $\alpha \sim (10^{-8} -10^{-5}) \rm s^{-1}$ and $\alpha \sim (10^{-10} - 10^{-7}) \rm s^{-1} $ is possible to be detected for a binary with total mass $5 M_{\odot}, q=10$ when observed in CE and ET respectively.

\section{Conclusion}
In this work, we explored the impact of a tertiary body on the GW emitted by a binary, focusing specifically on stellar-mass binary black holes inspiraling in the vicinity of a supermassive black hole (SMBH). The SMBH induces a CoM acceleration of the BBH which imprints a characteristic dephasing in the GW phase. Following \cite{Lazarow:2024gdn}, we consider the contributions of  CoM acceleration (approximated to the first order) to the PN expansion of the TaylorF2 waveform model. As the CoM acceleration parameter $\alpha$ gets larger, the higher-order terms become important and so we get the maximum allowed value of $\alpha$, i.e. $\alpha_{max}$ within which the waveform is valid. The $\alpha_{max}$ depends on the binary parameters and the detector’s lower cutoff frequency (see Fig.~\ref{fig:alpha_range}). Because the CoM acceleration dominates in low frequency regime, we concentrate on third-generation detectors CE and ET, whose improved low-frequency sensitivity makes them particularly well suited for this study.

We observe that for a low-mass and highly asymmetric BBH ($M=5,M_\odot$, $q=10$), the systematic bias induced by CoM acceleration can exceed the statistical uncertainty in both the chirp mass and the symmetric mass ratio when $\alpha \sim 10^{-10} \mathrm{s}^{-1}$. The bias is more pronounced for ET than for CE, consistent with ET’s slightly better low-frequency sensitivity. These results underscore that even modest CoM accelerations can significantly skew inference of key source parameters for lighter systems unless explicitly modeled.

On the other hand, including the effect of CoM acceleration allows us to get reasonable constraint on $\alpha$ from especially the lighter-mass BBHs. We show that a $M =5 M_{\odot}, q=10$ binary can constrain $\alpha = 5 \times 10^{-7} \rm s^{-1}$ with a $1\sigma$ bound of $9 \times 10^{-11} \rm s^{-1}$ if observed through ET at a distance of 500 Mpc. Throughout our analysis we have considered the binaries with aligned spin of $(0.2,0.1)$. Changing the spin of the binary is not expected to change the overall trends observed but will affect the absolute values of the biases for different binary configurations.

Thus, our study demonstrates that environmental effects like the induced CoM acceleration on a binary due to a  nearby SMBH can become detectable through third-generation GW observations. Even relatively small accelerations, $\alpha=\mathcal{O}(10^{-7})\mathrm{s}^{-1}$, can be constrained at the level of $\sim 10^{-9}\mathrm{s}^{-1}$ with CE and $\sim 10^{-10}\mathrm{s}^{-1}$ with ET, opening a pathway to probing SMBH neighborhoods and strengthening the case for dynamical formation channels in dense environments such as AGN disks.

\section{Acknowledgments}
The authors thank K. G. Arun for useful comments and suggestions. P. D. R. acknowledge the support from the Infosys Foundation. P. D. R. also acknowledges support from the National Science Foundation (NSF) via NSF Award No. PHY-2409372. S G acknowledges support from IIT Guwahati via the Institute Postdoctoral Fellowship (IPDF). S G also acknowledges the support from IACS via the Research Associate-I fellowship.

\bibliographystyle{apsrev}
\bibliography{ref}	
\end{document}